\documentclass[twocolumn,aps,superscriptaddress]{revtex4}
\usepackage{amssymb}
\input psfig

\def\IGNORE#1{}
\def\Mcal#1{{\mathcal{#1}}}
\def\Mbf#1{{\mathbf{#1}}}
\def\CHEM#1{${\mathsf{#1}}$}

\begin{document}

\title{Small Cycles in Small Worlds}

\author{Petra M. Gleiss}
\affiliation{Institut f\"ur Theoretische Chemie, 
 Universit\"at Wien
 W{\"a}hringerstra{\ss}e 17, A-1090 Wien,  Austria
}

\author{Peter F. Stadler$^{1,}$}
\affiliation{The Santa Fe Institute, 1399 Hyde Park Rd.,
 Santa Fe NM 87501, USA}

\author{Andreas Wagner$^{2,}$}
\affiliation{Dept.\ of Biology, University of New Mexico,
 167A Castetter Hall, Albuquerque, NM-817131-1091,
}

\author{David A. Fell}
\affiliation{School of Biological \& Molecular Sciences
 Oxford Brookes University, Oxford OX3 OBP, U.K.
}


\begin{abstract}
We characterize the distributions of short cycles in a large metabolic
network previously shown to have small world characteristics and a power
law degree distribution. Compared with three classes of random networks,
including Erd{\H o}s-R{\'e}nyi random graphs and synthetic small world
networks of the same connectivity, the metabolic network has a particularly
large number of triangles and a deficit in large cycles. Short cycles
reduce the length of detours when a connection is clipped, so we propose
that long cycles in metabolism may have been selected against in order to
shorten transition times and reduce the likelihood of oscillations in
response to external perturbations.
\end{abstract}


\maketitle

Systems as diverse as the Western US power grid, metabolic networks of a
cell, or the World Wide Web are well described as graphs with
characteristic topology. Small world networks have received considerable
attention since the seminal paper by Watts and Strogatz \cite{Watts:98}.

Most of the existing literature discusses small world networks in terms of
the average path length between two vertices \cite{Newman:00a} or of the
network's clustering coefficient \cite{Herzel:98,Barrat:00a} which measures
how close the neighborhood of a each vertex comes on average to being a
complete subgraph (clique) \cite{Watts:98}. Barab{\'a}si {\it et al.}
\cite{Barabasi:99a,Barabasi:99b} focussed on the degree distributions,
finding a power law in a suite of real world examples including the world wide web or
the US power-grid. Recent work on the spread of epidemics on a small world
network \cite{Pandit:99} emphasizes the importance of ``far-reaching''
edges. The idea is that clipping a far edge will force a (relatively) long
detour in the network. Hence it is these edges that are responsible for the
small diameter of the graph $G$.

Let us look at detours in graphs in more systematic way. Throughout this
paper we will represent a network as a simple (unweighted, undirected)
graph $G(V,E)$ with vertex set $V$ and edge set $E$. A {\em cycle} in $G$
is a closed path which meets each of its vertices and edges exactly
once. The length of a cycle $C$, i.e., the number of its vertices or edges,
is denoted by $|C|$.  With each edge $e\in E$ we can associate the set
$\Mcal{S}(e)$ containing the shortest cycles in $G$ that go through $e$.
It is easily verified that a far edge in the sense of \cite{Pandit:99} is
an edge that is not contained in a triangle. In other words, $e$ is a far
edge if and only if $\Mcal{S}(e)$ does not contain a triangle.  The cycles
$C\in\Mcal{S}(e)$ determine the shortest detours (which have length
$|C|-1$) when $e$ is removed from the graph.

\begin{figure}
\par
\centerline{\psfig{width=0.23\textwidth,file=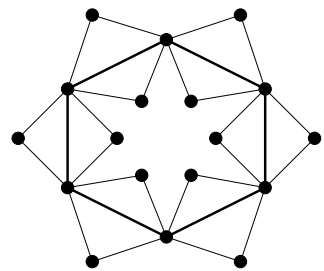}}
\par
\caption{$\Mcal{S}(G)$ consists of the twelve triangle only. The hexagon
(bold edges), however, is obviously crucial for the network structure.}
\label{Fig:SZ}
\end{figure}

It seems natural to consider the set $\Mcal{S}(G)=\bigcup_{e\in
E}\Mcal{S}(e)$ of shortest cycles of all edges in $G$ and to study e.g.\
their length distribution.  However, as the example in Fig.\ref{Fig:SZ}
shows, the shortest cycles $\Mcal{S}(G)$ do not convey the complete
information about the graph. Additional cycles appear to be relevant, such
as the hexagon in Figure~\ref{Fig:SZ}. In order to extend $\Mcal{S}(G)$ to
a more complete collection of cycles we need some more information on the
cycle structure of graphs. Recall that the set of all subsets of $E$ forms
an $|E|$-dimensional vector space over $\{0,1\}$ (with addition and
multiplication modulo $2$). Vector addition in this \emph{edge space} is
given by symmetric difference $X\oplus Y=(X\cup Y)\setminus(X\cap Y)$.  The
\emph{cycle space} $\Mcal{C}$ consisting of all cycles and edge-disjoint
unions of cycles in $G$ is a particularly important subspace of the edge
space \cite{Chen:71}. Its dimension is the \emph{cyclomatic number}
$\nu(G)=\vert E\vert - \vert V\vert +c(G)$, where $c(G)$ is the number of
connected components of $G$.  The length $\ell(\Mcal{B})$ of a basis
$\Mcal{B}$ of the cycle space (\emph{cycle basis} for short) is the sum of
the lengths of its cycles: $\ell(\Mcal{B}) = \sum_{C\in\Mcal{B}}\vert
C\vert$.  A \emph{minimum cycle basis} (MCB) is a cycle basis with minimum
length. MCBs have the property that their longest cycle is at most as long
as the longest cycle of any basis of $\Mcal{C}$ \cite{Chickering:95}. A MCB
therefore contains the salient information about the cycle structure of a
graph in its most compressed form.  Most graphs, however, do not have a
unique MCB. On the other hand, the distribution of cycle lengths is the
same in all MCBs of a given graph \cite{Stepanec:64}. The way to avoid
ambiguities is to consider the union of all minimum cycles bases, also
known as the set $\Mcal{R}(G)$ of \emph{ relevant cycles}. The term
``relevant'' is justified by two important properties of $\Mcal{R}(G)$: (i)
a cycle is relevant if and only if it cannot be written as an $\oplus$-sum
of shorter cycles \cite{Vismara:97}, and (ii) the shortest cycles through
an edge are relevant, i.e., $\Mcal{S}(G)\subseteq\Mcal{R}(G)$
\cite{Stepanec:64,Zykov:69}. Consequently, the composition of $\Mcal{R}(G)$
in terms of number and length distribution of cycles is an important
characteristic of a graph. The numerical studies below make use of
Vismara's \cite{Vismara:97} algorithm for computing $\Mcal{R}(G)$, which is
based on Horton's MCB algorithm \cite{Horton:87}.

The most common model of graph evolution, introduced by Erd{\H o}s and
R{\'e}nyi \cite{Erdoes:60}, assumes a fixed number $n=|V|$ of vertices and
assigns edges independently with a certain probability $p$
\cite{Bollobas:85}. In many cases ER random graphs turn out the be quite
different from a network of interest. The Watts-Strogatz \cite{Watts:98}
model of small world networks starts with a deterministic graph, usually a
circular arrangement of vertices in which each vertex is connected to $k$
nearest neighbors on each side. Then edges are ``rewired'' (in the original
version) or added \cite{Newman:99a,Newman:00a} with probability $p$. We shall consider
the latter model for $k=1$, denoted SW1 below, which corresponds to adding
random edges to a Hamiltonian cycle. Both ER and SW1 graphs exhibit an
approximately Gaussian degree distribution.

In many real networks, however, the degree distribution follows a power
law.  Barab{\'a}si {\it et al.}\ \cite{Barabasi:99a,Barabasi:99b} show that
the scale invariant behavior of the degree distributions can be explained
in terms of simple graph evolution model (AB model): Starting from a small
core graph, at each time step a vertex is added together with $m$ edges
that are connected to each previously present vertex $k$ with probability
$\Pi(k)=d(k)/\sum_j d(j)$, where $d(j)$ is the degree of vertex $j$. In
this contribution we will focus mostly on the AB model instead of Watt's
original construction, because we will apply the analysis of the cycle
structure to an empirical network for which a power-law like degree
distribution has been established. This network is the system of all
chemical reactions required for the synthesis of small-molecule building
blocks and energy in the bacterium {\it Escherichia coli}. Its structure
described in ref.\ \cite{Wagner:00a}. Such chemical reaction networks are
often referred to as metabolic networks.

\begin{figure}
\par
\centerline{\psfig{width=0.40\textwidth,file=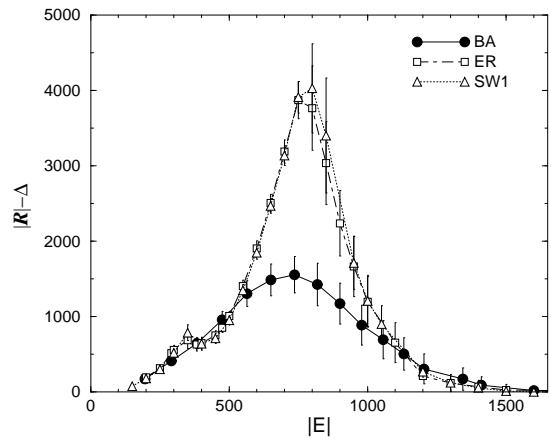}}
\par
\caption{Relevant non-triangles in ER ($\square$), SW1 ($\triangle$), and
AB ($\bullet$) random graphs with
$n=100$.}
\label{Fig:R100D}
\end{figure}

It is clear that all triangles in a graph are relevant, since a triangle is
for sure a shortest cycle through each edge. Hence
$|\Mcal{R}(G)|\ge\Delta$, where $\Delta$ denotes the number of triangles in
$G$.  We expect
$\langle\Delta\rangle_{\mathrm{ER}}=\left({n}\atop{3}\right)p^3$ triangles
in an ER random graph with edge-drawing probability $p$.  For the SW1
graphs we obtain a similar expression:
\begin{equation}
  \langle\Delta\rangle_{\mathrm{SW1}} =
  n p + n(n-4) p^2 + \frac{1}{6}n(n^2-9n+20) p^3 \,.
\label{eq:SW1tri}
\end{equation}
The MCB will therefore consist almost exclusively of triangles if
$\Delta\gg\nu(G)$. The average vertex degree is $d=2|E|/n=p(n-1)$ for ER
and $d=2+p(n-3)$ for SW1, resp. Assuming that $n$ is large we expect to
find only triangles in $\mathcal{R}(G)$ for $d\gg\sqrt{3n}$. Numerical
simulations show that this is indeed the case, Fig~\ref{Fig:R100D}. In this
regime, we have $|\Mcal{R}(G)|\sim d^3/6$, and the graph contains no far
edges. Not surprisingly, there is little difference between SW1 and ER
random graphs for large $n$. 

Since the AB model is based on a fixed vertex degree $d$, it should be
compared to random graph models with given vertex degree $d$, not with
given edge drawing probabilities $p$.  We have an asymptotically constant
number of triangles for both ER and SW1: $\Delta_{\mathrm{ER}}\to d^3/6$
and $\Delta_{\mathrm{SW1}}\to d^3/6-d+2/3$, resp. Note that as a
consequence the clustering coefficient vanishes asymptotically.  In SW
networks with {\em a priori} connectivity $k>1$ we find of course a number
of triangles that grows at least linearly with $n$, since the initial
($p=0$) networks already contains $(k-1)n$ triangles. The clustering
coefficient stays finite for large $n$ \cite{Watts:99}.

\begin{figure}
\par
\centerline{\psfig{width=0.40\textwidth,file=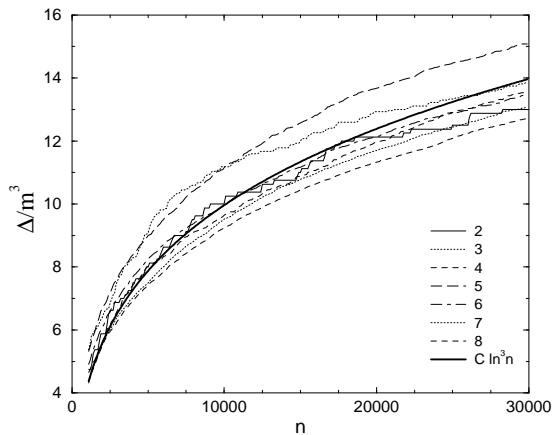}}
\par
\caption{Triangles in AB models with different values of $m$.}
\label{Fig:BATri}
\end{figure}

The large vertex degree of the ``early'' vertices in the AB model suggests
that there should be many more triangles than in ER or SW1 models. The
expected degree of vertex $s$ at ``time'' $t$ is known
\cite{Dorogovtsev:00}: $d(s|t) = m[\sqrt{t/s}-1]$. The probability of an
edge between $s$ and $t$, $t>s$, is therefore $p_{st} = m
d(s|t-1)/2(t-1)m$, where $2(t-1)m$ is the sum of the vertex degrees at
``time'' $t-1$. Thus $\langle \Delta\rangle = \sum_{r<s<t}
p_{rs}p_{st}p_{rt}$. This can be approximated by
\begin{eqnarray}
\langle\Delta\rangle &\approx&
\frac{m^3}{8}\!\int_{1<r<s<t}^n \!\!\!\!\!\!\!\!\!\!\!\!\!(1/st^2)
	     \left(\sqrt{\frac{s}{r}}-1\right)\!
             \left(\sqrt{\frac{t}{r}}-1\right)\!
	     \left(\sqrt{\frac{t}{s}}-1\right) \nonumber\\
&\sim& C m^3 \ln^3 n  + \Mcal{O}(\ln^2 n)
\label{eq:BATri}
\end{eqnarray}
Fig.~\ref{Fig:BATri} shows $\Delta$ for typical AB-random graphs with
$m=2,\dots,8$ as a function of ``time'. The behavior of $\Delta$ in
a individual growing network is well represented by equ.(\ref{eq:BATri}).

The number $|\Mcal{R}|-\Delta$ of non-trivial relevant cycles has its
maximum around $|E|\approx 0.74n^{3/2}$ independent of the model. The
scaling of their number is consistent with $|\Mcal{R}|-\Delta\sim C
n^{5/2}$, where the constant $C\approx 0.036$ is the same for ER and SW1
random graphs and $C\approx 0.016$ for the AB models. For small vertex
degrees, $d\ll|V|^{1/2}$ we find $\Mcal{R}(G)\approx \nu(G)$, i.e., the MCB
is (almost) unique.

\begin{figure}
\par
\centerline{\psfig{width=0.40\textwidth,file=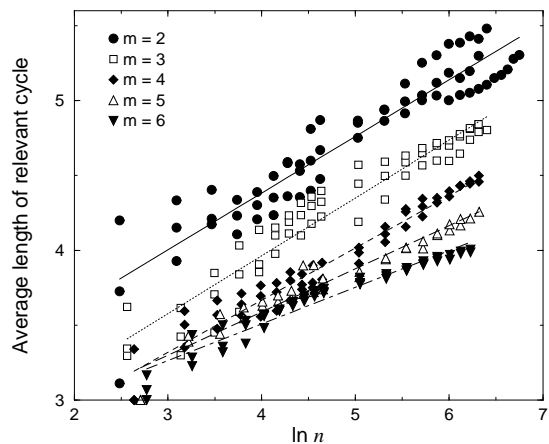}}
\par
\caption{Mean length of a relevant cycle in AB networks.}
\label{Fig:Bmeanl}
\end{figure}

The cyclomatic number of a AB random graph is $\nu(G)\sim (m/2-1)n$; Hence
eq.(\ref{eq:BATri}) implies that almost all relevant cycles must be long. 
Fig.~\ref{Fig:Bmeanl} shows that the average length of a relevant
cycle grows logarithmically with $n$. Not surprisingly, the slopes decrease
with $m$.

Let us now turn to an example of metabolic networks. Because it is germane
to their functional analysis, we first point out a nexus between graph
representations of metabolic network, and metabolic flux analysis (MFA),
the most generic framework to analyze the biological function of metabolic
networks.  A graph representation of metabolic networks was introduced as a
{\em substrate graph} $\Sigma$ in \cite{Wagner:00a}.  Its vertices are the
molecular compounds (substrates); two substrates $k$ and $l$ are adjacent
in $\Sigma$ if they participate in the same reaction $r$. Substrate graphs
are undirected because directed graphs would not properly represent the
propagation of perturbations: even for irreversible reactions the product
concentration may affect the the reaction rate, for instance by product
occupancy of the enzyme's active site; this in turn affects the substrate
concentration. Thus, perturbations may travel backwards even from
irreversible reactions. A similar argument for considering undirected
graphs can be derived from metabolic control theory \cite{Sen:91}.

The key ingredient of MFA is the {\em stoichiometric matrix} $\Mbf{S}$. Its
entries are the stoichiometric coefficients $s_{kr}$, i.e., the number of
molecules of species $k$ produced ($s_{kr}>0$) or consumed ($s_{kr}<0$) in
each reaction $r$. Reversible reactions are entered as two separate
reactions in most references. In general, additional ``pseudo-reactions''
are added to describe the interface of the metabolic reaction network with
its environment.  Stationary flux vectors $\vec f$ in the network satisfy
$\Mbf{S} f = \vec o$ and $f_r\ge0$ for each reaction $r$, see e.g.\
\cite{Clarke:88,Heinrich:96,Fell:97,Schilling:00,Edwards:00a,SchusterS:00a}.
It is not hard to see that if all reactions are mono-molecular, then
$\Mbf{S}$ is the incidence matrix of a directed graph; The stationary flux
vectors span the cycle space of this graph. The close connection between
the cycle space of a directed graph and its underlying undirected graph
\cite{Bollobas:98} allows us to use the relevant cycles of the substrate
graph $\Sigma$ to describe the structure of the metabolic network in a way
complementary to that provided by MFA.

For our analysis of metabolic graphs, we use the substrate graph of the
{\tt Ecoli1} core metabolism, a set of chemical reactions representing the
central routes of energy metabolism and small-molecule building block
synthesis. Similar to \cite{Wagner:00a}, we omit the following substrates
from the graph: \CHEM{CO_2}, \CHEM{NH_3}, \CHEM{SO_4},
\CHEM{AMP},\CHEM{ADP}, and \CHEM{ATP}, their deoxy-derivatives, both the
oxidized and reduced form of thioredoxine, organic phosphate and
pyrophosphate. The resulting graph has $n=272$ vertices and $|E|=652$
edges. It is analyzed below.

\begin{table}[t]
\caption{Cycle Structure of Metabolic Networks.}
\begin{ruledtabular}
\begin{tabular}{llrrrrrrrr}
Model        & $|C|$      &   3 &   4 &   5 &   6 &   7 &  8  &  9  & $\sum$\\
\hline 
{\tt Ecoli1} & MCB        & 282 &  51 &  19 &  20 &   3 &  5  &  1  & 381 \\
             & $\Mcal{R}$ & 379 & 114 &  90 &  83 &   5 & 36  & 16  & 723 \\
	     & $\Mcal{S}$ & 379 &  56 &  24 &  42 &   2 & 14  & 16  & 533 \\
\hline
AB           & MCB        &  78 & 158 & 124 &  20 & 0.4 & 0.01 & 0  & 380 \\
             & $\Mcal{R}$ &  81 & 285 & 527 & 161 & 5.5 & 0.4  & 0  &1060 \\
	     & $\Mcal{S}$ &  81 & 273 & 414 & 144 & 5.5 & 0.4  & 0  & 918 \\
\hline
ER           & MCB        &  18 &  58 & 163 & 131 &  11 & 0.4 & 0   & 381 \\
	     & $\Mcal{R}$ &  18 &  61 & 212 & 528 &  82 & 3.2 & 0   & 904 \\
	     & $\Mcal{S}$ &  18 &  61 & 205 & 311 &  68 & 3.2 & 0   & 666 \\
\hline
SW1          & MCB  	  &  15 &  46 & 131 & 167 &  21 & 1.1 & 0.03& 381 \\
 	     & $\Mcal{R}$ &  15 &  48 & 157 & 427 & 151 & 7.1 & 0.2 & 805 \\
 	     & $\Mcal{S}$ &  15 &  48 & 155 & 301 & 108 & 6.5 & 0.2 & 634 \\
\end{tabular}
\end{ruledtabular}
\label{tab:network}
\end{table}

Table~\ref{tab:network} shows that the three random models AB, SW1, and ER
agree at least qualitatively with each other. The AB random graphs exhibit
a much broader distribution of cycle sizes (not shown) than the ER and SW1
models. As a consequence, the average cycle numbers for ER and SW1 have
statistical uncertainty of about 2\%, while the uncertainty of the AB
values is 5 to 10 times higher. Note that ER and SW1 have a similar number
of relevant cycles, but the cycles are slightly longer in SW1.  Two
features distinguish the metabolic network {\tt Ecoli1} from all random
networks: (1) The number $\Delta$ of triangles is almost 10 times larger
than expected. This can be explained at least in part as a consequence of
the substrate graph representation: multi-molecular reactions translate to
cliques and hence a large number of triangles. The ratio $282/379\approx
0.744$ indicates that in fact almost all triangles are contained in
4-cliques, since in each 4-clique we have three triangles that belong to a
particular MCB, while the fourth face of the tetrahedron is their
$\oplus$-sum \cite{Gleiss:00a}. (2) There is a much smaller number of
relevant pentagons and hexagons, which results in an overall somewhat
reduced number of relevant cycles: $723$ compared to about $1060$ (AB),
$904$ (ER), and $805$ (SW1).

Strictly speaking, we do not know the biological significance of this
relative paucity of longer cycles. However, we would like to venture a
speculation. Organisms are constantly exposed to environmental fluctuations
requiring transitions in metabolic states. That is, a metabolic network
needs to produce different outputs depending on the environment.
Environments may vary rapidly, requiring rapid transition between metabolic
states. Possibly, networks with long cycles have longer transition times,
because environmental perturbations may lead to prolonged oscillations in
such networks. The dynamical system representation of metabolic networks
required to test this idea rigorously lies beyond the scope of this
article.

\begin{figure}
\par
\centerline{\psfig{width=0.40\textwidth,file=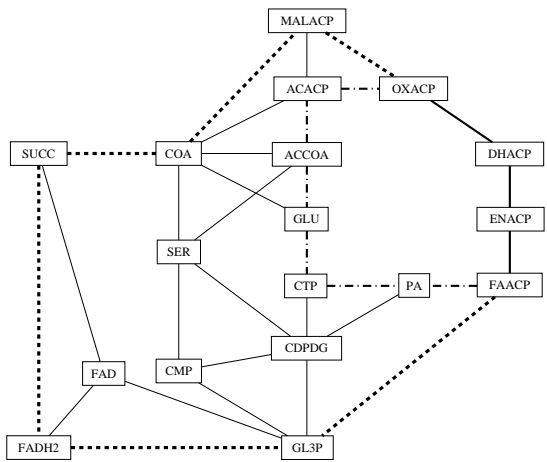}}
\par\caption{The subgraph of {\tt Ecoli} spanned by the relevant cycles of
length 9. Two of these long cycles are highlighted. The edges shown in bold
are part of each of the 16 relevant 9-cycles.}
\label{fig:9cycle}
\end{figure}

The longest relevant cycles in a metabolic networks are of particular
interest since they reflect parts of the network that cannot easily be
replaced by alternative routes. In Fig.~\ref{fig:9cycle} we show the
largest such cycle in {\tt Ecoli1}. We emphasize that the cycles in our
analysis represent routes for transmission of perturbations, but not
necessarily of mass, as it is commonly considered in MFA. This is apparent
from Fig.\ref{fig:9cycle} , which does not correspond to a pathway from a
biochemical chart, but links serval pathways together.



\end{document}